\documentclass[aps,twocolumn,pra,superscriptaddress,showpacs,tightenlines]{revtex4-2}
\usepackage{amssymb}
\usepackage{amsmath}
\usepackage{graphicx}
\usepackage{txfonts}
\usepackage{color}
\usepackage{amsfonts}
\usepackage[colorlinks,urlcolor=black,linkcolor=blue,citecolor=blue]{hyperref}

\begin{document}
\title{Dynamic sensitivity of quantum Rabi model with quantum criticality}
\author{Ying Hu}
\affiliation{Key Laboratory of Low-Dimensional Quantum Structures and Quantum Control
of Ministry of Education, Key Laboratory for Matter Microstructure
and Function of Hunan Province, Department of Physics and Synergetic
Innovation Center for Quantum Effects and Applications, Hunan Normal
University, Changsha 410081, China}
\author{Jian Huang}
\affiliation{Key Laboratory of Low-Dimensional Quantum Structures and Quantum Control
of Ministry of Education, Key Laboratory for Matter Microstructure
and Function of Hunan Province, Department of Physics and Synergetic
Innovation Center for Quantum Effects and Applications, Hunan Normal
University, Changsha 410081, China}
\author{Jin-Feng Huang}
\email{jfhuang@hunnu.edu.cn}

\affiliation{Key Laboratory of Low-Dimensional Quantum Structures and Quantum Control
of Ministry of Education, Key Laboratory for Matter Microstructure
and Function of Hunan Province, Department of Physics and Synergetic
Innovation Center for Quantum Effects and Applications, Hunan Normal
University, Changsha 410081, China}
\author{Qiong-Tao Xie}
\email{qiongtaoxie@yahoo.com}

\affiliation{College of Physics and Electronic Engineering, Hainan Normal University,
Haikou 571158, China}
\author{Jie-Qiao Liao}
\email{jqliao@hunnu.edu.cn}

\affiliation{Key Laboratory of Low-Dimensional Quantum Structures and Quantum Control
of Ministry of Education, Key Laboratory for Matter Microstructure
and Function of Hunan Province, Department of Physics and Synergetic
Innovation Center for Quantum Effects and Applications, Hunan Normal
University, Changsha 410081, China}
\date{\today}

\begin{abstract}
We study the dynamic sensitivity of the quantum Rabi model, which
exhibits quantum criticality in the finite-component-system case.
This dynamic sensitivity can be detected by introducing an auxiliary
two-level atom far-off-resonantly coupled to the cavity field of the
quantum Rabi model. We find that when the quantum Rabi model goes
through the critical point, the auxiliary atom experiences a sudden
decoherence, which can be characterised by a sharp decay of the Loschmidt
echo. Our scheme will provide a reliable way to observe quantum phase
transition in ultrastrongly coupled quantum systems. 
\end{abstract}
\maketitle
 
\section{INTRODUCTION}

Quantum phase transition (QPT) \cite{sachdev_2011}, as a fundamental
phenomenon in quantum physics, is characterized by the sudden change
of the ground states of quantum systems, induced by the change of
system parameters. In general, a quantum system occurred phase transition
needs to reach a thermodynamic limit, i.e., the components of the
system should attain infinity~\cite{HEPP1973360,PhysRevA.7.831,PhysRevLett.90.044101,PhysRevE.67.066203,PhysRevB.81.121105_2010}.
Since the infinite-component systems are composed of numerous degrees
of freedom, it will take a long time to prepare the initial state
of the systems, and the systems are easily affected by their environments.
Consequently, an interesting question is whether quantum phase transition
can take place in a finite-component system. It has recently been
shown that quantum phase transition can take place in simple systems~\cite{PhysRevA.54.R4657,Bishop_2001,PhysRevB.69.113203,PhysRevA.70.022303,PhysRevA.81.042311,PhysRevA.82.025802,PhysRevA.82.053841,PhysRevB.81.121105_2010,PhysRevA.87.013826,PhysRevLett.115.180404,PhysRevA.92.053823_2015,PhysRevLett.117.123602,PhysRevA.94.023835,PhysRevLett.119.220601,PhysRevA.95.013819,PhysRevA.95.043823}.
An advantage of this kind of systems is that the systems have less
degrees of freedom, and hence those previous mentioned difficulties
in infinite-component systems can be improved~\cite{PhysRevLett.124.120504}.

Quantum Rabi model (QRM), as a typical finite-component quantum system,
describes the interaction between a single two-level atom (qubit)
and a single bosonic mode. As one of the most fundamental models in
quantum optics, the QRM has attracted much attention from the communities
of quantum physics, quantum information, and especially ultrastrong
couplings~\cite{NatureREVPHY2019,RevModPhys.91.025005,NaturePhysics6772}.
In QRM, it has been shown that the quantum criticality exists in a
limit case, in which the ratio $\eta=\omega_{0}/\omega_{c}$ of the
qubit frequency $\omega_{0}$ to the bosonic-mode frequency $\omega_{c}$
tends to infinity. It has also been recognized that the quantum critical
position of the QRM depends on the frequency of the bosonic mode,
and that the coupling strength needs to enter the ultrastrong-coupling
regime~\cite{NatureREVPHY2019,RevModPhys.91.025005,NaturePhysics6772}.
Currently, with the improvement of the experimental conditions, the
ultrastrong couplings even deep-strong couplings have been realized
in various physical systems, such as superconducting quantum circuits~\cite{NaturePhysics6772,PhysRevLett.105.237001,PhysRevLett.105.060503,Forn2017Ultrastrong,2016Superconducting}
and semiconductor quantum wells~\cite{PhysRevB.79.201303,Nature7235178,PhysRevLett.105.196402}.
In the ultrastrong-coupling regime, the coupling strength is comparable
to the frequencies of the bosonic mode and the two-level atom. Furthermore,
the quantum criticality in finite-component quantum systems has been
expected and experimentally demonstrated in trapped-ion systems~\cite{Nat.Commun.2.377_2011,PhysRevLett.118.073001_2017,PhysRevX.8.021027_2018,PhysRevA.97.042317_2018,PhysRevA.100.022513_2019}.
All these advances motivate the experimental studies of quantum criticality
in various finite-component quantum systems, and hence how to observe
quantum criticality in realistic finite-component quantum systems
becomes an interesting task.

In this paper, we propose to show the dynamic sensitivity caused by
quantum criticality in the QRM by introducing an auxiliary atom coupled
to the cavity field of the QRM. Here, the auxiliary atom plays two
important roles in this system. The first role is a trigger, which
is used to stir up the quantum criticality in this system. The second
role is a sensor to detect the quantum critical behavior in the QRM.
Here, the decoherence of the auxiliary atom can reflect the dynamic
sensitivity of the QRM, and we use the Loschmidt echo (LE) to measure
the decoherence of the auxiliary atom~\cite{PhysRevLett.96.140604,PhysRevA.80.063829}.
In the short-time limit, the LE can be simplified to an exponential
function of the photon number variance in the ground state of the
QRM. As a result, we calculate the ground state of QRM in both the
normal and superradiance phases, and compare the analytical ground
state in the infinite $\eta$ case with the numerical ground state
in the finite $\eta$ case~\cite{PhysRevLett.115.180404}. We also
analyze the LE of the auxiliary atom and find a dynamic sensitivity
around the critical point of the QRM. This feature provides a signature
to characterize the quantum criticality in this system.

The rest of this paper is organized as follows. In Sec.~\ref{section2},
we introduce the QRM and construct conditional Rabi interactions by
introducing an auxiliary two-level atom far-off-resonantly coupled
to the cavity field in the QRM. In Sec.~\ref{section3}, we present
the analytical result for the LE of the auxiliary atom. We also calculate
the ground state of the critical QRM when it works in both the normal
and the superradiance phases at both finite and infinite $\eta$.
In Sec.~\ref{section4}, we exhibit the dependence of the LE on the
system parameters and analyze the dynamic sensitivity of the QRM.
Finally, a brief summary of this paper is presented in Sec.~\ref{section5}.

\section{MODEL AND HAMILTONIAN}

\label{section2}

We consider the quantum Rabi model, which is composed of a single-mode
cavity field coupled to a two-level atom. The Hamiltonian of the QRM
reads ($\hbar=1$)~\cite{NatureREVPHY2019,RevModPhys.91.025005}
\begin{equation}
\hat{H}_{\mathrm{Rabi}}=\omega_{c}\hat{a}^{\dagger}\hat{a}+\dfrac{\omega_{0}}{2}\hat{\sigma}_{z}-g\hat{\sigma}_{x}(\hat{a}+\hat{a}^{\dagger}),\label{eq:1}
\end{equation}
where $\hat{a}\ (\hat{a}^{\dagger})$ is the annihilation (creation)
operator of the cavity field with resonance frequency $\omega_{c}$.
The two-level atom has the ground state $\vert g\rangle$ and excited
state $\vert e\rangle$ with transition frequency $\omega_{0}$, and
it is described by the Pauli operators $\hat{\sigma}_{x}\equiv\vert e\rangle\langle g\vert+\vert g\rangle\langle e\vert$,
$\hat{\sigma}_{y}\equiv i(\vert g\rangle\langle e\vert-\vert e\rangle\langle g\vert)$,
and $\hat{\sigma}_{z}\equiv\vert e\rangle\langle e\vert-\vert g\rangle\langle g\vert$.
The parameter $g$ denotes the coupling strength between the cavity
field and the atom. In QRM, the parity operator $\hat{\Pi}=\exp\{i\pi[\hat{a}^{\dagger}\hat{a}+(1+\hat{\sigma}_{z})/2]\}$
is a conserved quantity based on the commutative relation $[\hat{\Pi},\hat{H}_{\mathrm{Rabi}}]=0$,
then the Hilbert space of the QRM can be divided into two subspaces
with odd and even parities. It is known that the QRM has a $Z_{2}$
parity symmetry and it is integrable~\cite{PhysRevLett.105.263603_2010,PhysRevLett.107.100401,PhysRevA.87.023835_2013}.
The analytic eigensystem is determined by a transcendental equation
given in Refs.~\cite{PhysRevLett.107.100401,PhysRevA.86.023822,PhysRevX.4.021046,PhysRevA.90.063839}.
Due to the lack of closed-form solution, several methods for approximatively
solving the QRM have also been proposed~\cite{PhysRevB.72.195410,PhysRevLett.99.173601,PhysRevA.83.065802,PhysRevA.86.015803,Zhong_2013,PhysRevA.92.053823_2015,Liu_2015,PhysRevA.91.053834,PhysRevA.94.063824}.

It has been found that a QPT takes place in the QRM at the critical
point $g=\sqrt{\omega_{c}\omega_{0}}/2$~\cite{PhysRevLett.115.180404}.
This feature motivates us to detect the dynamic sensitivity of quantum
criticality in the QRM by introducing an auxiliary atom $S$ coupled
to the cavity field of the QRM. The auxiliary atom and its interaction
with the cavity field are described by the Hamiltonian 
\begin{equation}
\hat{H}_{I}=\dfrac{\omega_{s}}{2}\hat{\sigma}_{z}^{(s)}-g_{s}(\hat{a}^{\dagger}\hat{\sigma}_{-}^{(s)}+\hat{\sigma}_{+}^{(s)}\hat{a}),\label{eq:2}
\end{equation}
where $\hat{\sigma}_{z}^{(s)}=\vert e\rangle_{\!s\,s\!}\langle e\vert-\vert g\rangle_{\!s\,s\!}\langle g\vert$
is the $z$-direction Pauli operator, $\hat{\sigma}_{+}^{(s)}=\vert e\rangle_{\!s\,s\!}\langle g\vert$
and $\hat{\sigma}_{-}^{(s)}=\vert g\rangle_{\!s\,s\!}\langle e\vert$
are the raising and lowing operators of the auxiliary atom, respectively.
$\omega_{s}$ is the transition frequency between the ground state
$\vert g\rangle_{\!s}$ and excited states $\vert e\rangle_{\!s}$
of the auxiliary atom. $g_{s}$ is the coupling strength between the
cavity field and the auxiliary atom. Note that here we consider the
case where the interaction between the auxiliary atom and the cavity
field works in the Jaynes-Cummings (JC) coupling regime~\cite{1443594}
and then the rotating-wave approximation has been made in Hamiltonian~(\ref{eq:2}).

We assume that the auxiliary two-level atom $S$ is far-off-resonantly
coupled with the single-mode cavity field, namely the detuning $\Delta_{s}\equiv\omega_{s}-\omega_{c}$
is much larger than the coupling strength $g_{s}\sqrt{n}$ with $n$
being the involved photon number. In the large-detuning regime, the
interaction between the auxiliary atom $S$ and the cavity field is
described by the dispersive JC model~\cite{Wolfgang_2001}. Therefore,
the Hamiltonian of the whole system including the QRM and the auxiliary
atom reads 
\begin{align}
\hat{H}_{\mathrm{eff}}= & \ \omega_{c}\hat{a}^{\dagger}\hat{a}+\frac{\omega_{0}}{2}\hat{\sigma}_{z}-g\hat{\sigma}_{x}(\hat{a}+\hat{a}^{\dagger})\nonumber \\
 & +\frac{1}{2}\omega_{s}\hat{\sigma}_{z}^{(s)}+\chi\hat{\sigma}_{z}^{(s)}\hat{a}^{\dagger}\hat{a}+\chi\hat{\sigma}_{+}^{(s)}\hat{\sigma}_{-}^{(s)},\label{eq:3}
\end{align}
where $\chi\equiv g_{s}^{2}/\Delta_{s}$ is the dispersive JC coupling
strength between the auxiliary atom and the cavity field. The dispersive
JC coupling describes a conditional frequency shift for the cavity
field. To clearly see the dynamic sensitivity of the finite-component
system response to the auxiliary atom, we rewrite Hamiltonian~(\ref{eq:3})
as the following form 
\begin{equation}
\hat{H}_{\mathrm{eff}}=\hat{H}_{e}\otimes\vert e\rangle_{\!s\,s\!}\langle e\vert+\hat{H}_{g}\otimes\vert g\rangle_{\!s\,s\!}\langle g\vert,\label{eq:4}
\end{equation}
where 
\begin{align}
\hat{H}_{e} & =\omega_{e}\hat{a}^{\dagger}\hat{a}+\frac{\omega_{0}}{2}\hat{\sigma}_{z}-g\hat{\sigma}_{x}(\hat{a}+\hat{a}^{\dagger})+\frac{\omega_{s}}{2}+\chi,\label{eq:5}\\
\hat{H}_{g} & =\omega_{g}\hat{a}^{\dagger}\hat{a}+\frac{\omega_{0}}{2}\hat{\sigma}_{z}-g\hat{\sigma}_{x}(\hat{a}+\hat{a}^{\dagger})-\frac{\omega_{s}}{2},\label{eq:6}
\end{align}
with the state-dependent cavity frequencies $\omega_{e}=\omega_{c}+\chi$
and $\omega_{g}=\omega_{c}-\chi$. Note that up to the constant terms,
both the two Hamiltonians in Eqs.~(\ref{eq:5}) and (\ref{eq:6})
describe the QRM with different cavity-field frequencies.

\section{QUANTUM CRITICAL EFFECT }

\label{section3} In this section, we derive the relation between
the LE and the photon number variance of the cavity field. We also
calculate the expression of the photon number varianance in the finite
and infinite $\eta$ cases when the QRM works in both the normal and
superradiance phases.

\subsection{Expression of the LE}

To study quantum critical effect in this system, we investigate the
dynamic evolution of the system, which is governed by Hamiltonian~(\ref{eq:4}).
To this end, we assume that the QRM is initially in its ground state
$\vert G\rangle$ and the auxiliary atom is in a superposed state
$\alpha\vert g\rangle_{\!s}+\beta\vert e\rangle_{\!s}$, where $\alpha$
and $\beta$ are the superposition coefficients, satisfying the normalization
condition $\vert\alpha\vert^{2}+\vert\beta\vert^{2}=1$. Corresponding
to the auxiliary atom in states $\vert g\rangle_{\!s}$ and $\vert e\rangle_{\!s}$,
the evolution of the QRM is governed by the Hamiltonians $\hat{H}_{g}$
and $\hat{H}_{e}$, respectively. Then, the state of the total system
at time $t$ becomes 
\begin{equation}
\left|\Psi(t)\right\rangle =\alpha\vert g\rangle_{\!s}\otimes\vert\Phi_{g}(t)\rangle+\beta\vert e\rangle_{\!s}\otimes\vert\Phi_{e}(t)\rangle,\label{eq:7}
\end{equation}
where $\vert\Phi_{g}(t)\rangle\equiv\mathrm{e}^{-i\hat{H}_{g}t}\vert G\rangle$
and $\vert\Phi_{e}(t)\rangle\equiv\mathrm{e}^{-i\hat{H}_{e}t}\vert G\rangle$.
The central task of this paper is to study the dynamic sensitivity
of the QRM with respect to the state of the auxiliary atom, which
could play the role of a sensor to detect the criticality of QRM.
To show this physical mechanism, we trace over the degrees of freedom
of the QRM, and obtain the reduced density matrix of the auxiliary
atom as 
\begin{equation}
\hat{\rho}_{s}(t)=\vert\alpha\vert^{2}\vert g\rangle_{\!s\,s\!}\langle g\vert+\vert\beta\vert^{2}\vert e\rangle_{\!s\,s\!}\langle e\vert+[D(t)\alpha^{\ast}\beta\vert e\rangle_{\!s\,s\!}\langle g\vert+\mathrm{H.c.}],\label{eq:8}
\end{equation}
where the decoherence factor $D(t)$ is defined by 
\begin{equation}
D(t)=\langle\Phi_{g}(t)\vert\Phi_{e}(t)\rangle=\langle G\vert\mathrm{e}^{i\hat{H}_{g}t}\mathrm{e}^{-i\hat{H}_{e}t}\vert G\rangle.\label{eq:9}
\end{equation}
To understand the dynamic sensitivity in this finite-component system,
we calculate the LE of the auxiliary atom by 
\begin{equation}
L(t)=\vert D(t)\vert^{2}=\vert\langle G\vert\mathrm{e}^{i\hat{H}_{g}t}\mathrm{e}^{-i\hat{H}_{e}t}\vert G\rangle\vert^{2}.\label{eq:10}
\end{equation}
In the short-time limit, the LE can be approximated as 
\begin{equation}
L(t)\approx\exp(-4\gamma\chi^{2}t^{2}),\label{eq:11}
\end{equation}
where $\gamma=\langle G\vert(\hat{a}^{\dagger}\hat{a})^{2}\vert G\rangle-\langle G\vert\hat{a}^{\dagger}\hat{a}\vert G\rangle^{2}$
is the photon number variance. Note that the operator average here
is taken over the ground state of the QRM. Equation~(\ref{eq:11})
shows that the decay rate of the LE depends on $t^{2}$ and the photon
number variance $\gamma$. To obtain the LE, we need to know the ground
states of the QRM working in both the normal and the superradiance
phases.

The QRM undergoes a quantum phase transition from the normal phase
to the superradiance phase by increasing the coupling strength crossing
the critical point $g_{c}=\sqrt{\omega_{c}\omega_{0}}/2$. When the
QRM goes through the critical point, the ground state of the QRM experience
a huge change. In this paper, we will exhibit some special features
around the critical point by calculating the photon number variance
$\gamma$ in these two phases. When $L(t)$ approaches zero, the QRM
will evolve into two orthogonal states $\vert\Phi_{g}(t)\rangle$
and $\vert\Phi_{e}(t)\rangle$. This feature could be used as a measurement
tool for detecting the state of the auxiliary atom.

\subsection{Photon number variance in the normal phase}

In this subsection, we calculate the photon number variance $\gamma$
in the ground state of the QRM working in the normal phase. Note that
the ground state of the QRM in the infinite and finite $\eta$ cases
have been calculated in Ref.~\cite{PhysRevLett.115.180404}. Here,
we present the calculation of the ground state for keeping the completeness
of this paper.

\subsubsection{The infinite $\eta$ case}

We first consider the infinite-frequency limit, i.e., the ratio $\eta=\omega_{0}/\omega_{c}$
of the atomic transition frequency $\omega_{0}$ over the cavity-field
frequency $\omega_{c}$ approaches infinity. In this case, the quantum
criticality has been analytically found in QRM~\cite{PhysRevLett.115.180404}.
By introducing the unitary transformation operator $\hat{U}_{\mathrm{np}}=\exp[i(g/\omega_{0})(\hat{a}+\hat{a}^{\dagger})\hat{\sigma}_{y}]$,
the Hamiltonian~(\ref{eq:1}) can be transformed to a decoupling
form corresponding to the spin subspaces $\mathcal{H}_{e}$ and $\mathcal{H}_{g}$.
Keeping the terms up to the second order of $g/\omega_{0}$, the transformed
Hamiltonian becomes~\cite{PhysRevLett.115.180404} 
\begin{align}
\hat{H}_{\mathrm{np}} & =\hat{U}_{\mathrm{np}}^{\dagger}\hat{H}_{\mathrm{Rabi}}\hat{U}_{\mathrm{np}}\nonumber \\
 & \approx\omega_{c}\hat{a}^{\dagger}\hat{a}+\frac{\omega_{c}\lambda^{2}}{4}(\hat{a}+\hat{a}^{\dagger})^{2}\hat{\sigma}_{z}+\frac{\omega_{0}}{2}\hat{\sigma}_{z}+\hat{\mathcal{O}}[(g/\omega_{0})^{2}],\label{eq:12}
\end{align}
where we introduce the dimensionless coupling strength $\lambda=2g/\sqrt{\omega_{0}\omega_{c}}$.
Hamiltonian~(\ref{eq:12}) can be diagonalized in terms of the squeezing
operator $\hat{S}(r_{\mathrm{np}})=\exp[r_{\mathrm{np}}(\hat{a}^{\dagger2}-\hat{a}^{2})/2]$
with $r_{\mathrm{np}}(\lambda)=-\frac{1}{4}\ln(1+\lambda^{2}\hat{\sigma}_{z})$.
The diagonalized Hamiltonian reads~\cite{PhysRevLett.115.180404}
\begin{equation}
\hat{H}_{\mathrm{np}}^{\mathrm{d}}=\hat{S}^{\dagger}(r_{\mathrm{np}})\hat{H}_{\mathrm{np}}\hat{S}(r_{\mathrm{np}})=\hat{\epsilon}_{\mathrm{np}}\hat{a}^{\dagger}\hat{a}+\hat{E}_{\mathrm{np}},\label{eq:13}
\end{equation}
where we introduce the conditional frequency $\hat{\epsilon}_{\mathrm{np}}=\omega_{c}\sqrt{1+\lambda^{2}\hat{\sigma}_{z}}$
and the spin-state dependent energy $\hat{E}_{\mathrm{np}}=(\hat{\epsilon}_{\mathrm{np}}-\omega_{c}+\omega_{0}\hat{\sigma}_{z})/2$.
By finding the minimum energy, the ground state of the diagonalized
Hamiltonian~(\ref{eq:13}) in the normal phase is $\text{\ensuremath{\vert}}0\rangle\ensuremath{\vert}g\rangle$.
To keep the Hamiltonian $\hat{H}_{\mathrm{np}}^{\mathrm{d}}$ in the
low spin subspace to be Hermitian, the coupling strength $g$ should
be smaller than $\sqrt{\omega_{0}\omega_{c}}/2$ such that $\lambda<\lambda_{c}=1$,
which defines the parameter space of the \textit{normal phase}. In
our model, the cavity frequency $\omega_{c}$ conditionally depends
on the states of the auxiliary atom, and hence the atom $S$ can be
used as a trigger to induce the criticality in the QRM.

Based on the above analyses, we know that the ground state of the
QRM is approximately expressed as~\cite{PhysRevLett.115.180404}
\begin{equation}
\vert\psi_{\mathrm{np}}^{G}(r_{\mathrm{np}})\rangle=\hat{U}_{\mathrm{np}}\hat{S}(r_{\mathrm{np}})\vert0\rangle\vert g\rangle,\label{eq:15}
\end{equation}
where the operators $\hat{U}_{\mathrm{np}}$ and $\hat{S}(r_{\mathrm{np}})$
have been defined before. In the ground state $\vert\psi_{\mathrm{np}}^{G}\rangle$,
the photon number variance in normal phase can be obtained as 
\begin{equation}
\gamma_{\mathrm{np}}=\frac{1}{2}\sinh^{2}(2r_{\mathrm{np}})+\frac{g^{2}}{\omega_{0}^{2}}\mathrm{e}^{-2r_{\mathrm{np}}}.\label{eq:16}
\end{equation}
In terms of Eqs.~(\ref{eq:11}) and~(\ref{eq:16}), the analytical
result of the LE in the normal phase can be obtained.

\subsubsection{The finite $\eta$ case}

The above discussions are valid in the infinite $\eta$ case. To beyond
this limit case, below we calculate the ground state of the QRM in
a large finite $\eta$ case. To this end, we perform a unitary transfomation
with the transformation operator
\begin{equation}
\hat{U}_{\mathrm{np}}^{\sigma}=\exp\left\{ i\left[\frac{g}{\omega_{0}}(\hat{a}+\hat{a}^{\dagger})-\frac{4g^{3}}{3\omega_{0}^{3}}(\hat{a}+\hat{a}^{\dagger})^{3}\right]\hat{\sigma}_{y}\right\} \label{eq:17}
\end{equation}
to the Hamiltonian $\hat{H}_{\mathrm{Rabi}}$~\cite{PhysRevLett.115.180404}.
Up to the fourth order of $g/\omega_{0}$, the transformed Hamiltonian
becomes~\cite{PhysRevLett.115.180404} 
\begin{align}
\hat{\tilde{H}}_{\mathrm{np}}^{\sigma}= & \ (\hat{U}_{\mathrm{np}}^{\sigma})^{\dagger}\hat{H}_{\mathrm{Rabi}}\hat{U}_{\mathrm{np}}^{\sigma}\nonumber \\
= & \ \omega_{c}\hat{a}^{\dagger}\hat{a}+\frac{g^{2}}{\omega_{0}}(\hat{a}+\hat{a}^{\dagger})^{2}\hat{\sigma}_{z}-\frac{g^{4}}{\omega_{0}^{3}}(\hat{a}+\hat{a}^{\dagger})^{4}\hat{\sigma}_{z}\nonumber \\
 & +\frac{\omega_{0}}{2}\hat{\sigma}_{z}+\frac{g^{2}\omega_{c}}{\omega_{0}^{3}}+\hat{\mathcal{O}}[(g/\omega_{0})^{4}].\label{eq:18}
\end{align}
By projecting the effective Hamiltonian~(\ref{eq:18}) into the low
spin subspace $\mathcal{H}_{g}$, we obtain 
\begin{align}
\hat{H}_{\mathrm{np}}^{\sigma}= & \ \langle g\vert\hat{\tilde{H}}_{\mathrm{np}}^{\sigma}\vert g\rangle\nonumber \\
= & \ \omega_{c}\hat{a}^{\dagger}\hat{a}-\frac{\omega_{c}\lambda^{2}}{4}(\hat{a}+\hat{a}^{\dagger})^{2}+\frac{\lambda^{4}\omega_{c}^{2}}{16\omega_{0}}(\hat{a}+\hat{a}^{\dagger})^{4}\nonumber \\
 & -\frac{\omega_{0}}{2}+\frac{\lambda^{2}\omega_{c}^{2}}{4\omega_{0}}.\label{eq:19}
\end{align}

To know the ground state of the Hamiltonian $\hat{H}_{\mathrm{np}}^{\sigma}$,
we adopt the variational method and assume a trial wave function $\vert\Psi_{\mathrm{np}}^{G}(s_{\mathrm{np}})\rangle=\hat{S}(s_{\mathrm{np}})\vert0\rangle\vert g\rangle$,
where $\hat{S}$ is a squeezing operator, and $s_{\mathrm{np}}$ is
the undetermined variational squeezing parameter~\cite{PhysRevLett.115.180404,PhysRevA.94.063824}.
The ground-state energy can be calculated as 
\begin{align}
E_{\mathrm{np}}^{G}(s_{\mathrm{np}})= & \ \omega_{c}\sinh^{2}s_{\mathrm{np}}-\frac{\omega_{c}\lambda^{2}}{4}\mathrm{e}^{2s_{\mathrm{np}}}+\frac{3\lambda^{4}\omega_{c}^{2}}{16\omega_{0}}\mathrm{e}^{4s_{\mathrm{np}}}\nonumber \\
 & -\frac{\omega_{0}}{2}+\frac{\lambda^{2}\omega_{c}^{2}}{4\omega_{0}}.\label{eq:19-1}
\end{align}
Here, the second-order derivative of $E_{\mathrm{np}}^{G}(s_{\mathrm{np}})$
with respect to $s_{\mathrm{np}}$ is positive, and then the minimum
energy can be obtained by the zero point of the first-order derivative~\cite{PhysRevLett.115.180404},
namely 
\begin{equation}
\frac{\mathrm{d}E_{\mathrm{np}}^{G}(s_{\mathrm{np}})}{\mathrm{d}s_{\mathrm{np}}}=\ \frac{\omega_{c}}{2\mathrm{e}^{2s_{\mathrm{np}}}}\left[\frac{3\lambda^{4}\mathrm{e}^{6s_{\mathrm{np}}}}{2\eta}+(1-\lambda^{2})\mathrm{e}^{4s_{\mathrm{np}}}-1\right]=0.\label{eq:21}
\end{equation}
By solving Eq.~(\ref{eq:21}), we obtain the only physical solution
as 
\begin{equation}
s_{\mathrm{np}}=\frac{1}{2}\ln\left\{ \mathrm{Re}\left[\frac{\sqrt[3]{A}}{9\lambda^{4}}+\frac{2(\lambda^{2}-1)\eta}{9\lambda^{4}}+\frac{4(\lambda^{2}-1)^{2}\eta^{2}}{9\lambda^{4}\sqrt[3]{A}}\right]\right\} ,\label{eq:21-1}
\end{equation}
where we introduce 
\begin{align}
A= & \ 9\sqrt{3}\sqrt{243\lambda^{16}\eta^{2}+(\lambda^{6}-3\lambda^{4}+3\lambda^{2}-1)16\lambda^{8}\eta^{4}}\nonumber \\
 & +243\lambda^{8}\eta+(\lambda^{6}-3\lambda^{4}+3\lambda^{2}-1)8\eta^{3}.
\end{align}
The ground state of QRM in the finite $\eta$ case can be expressed
as 
\begin{equation}
\vert\varphi_{\mathrm{np}}^{G}\rangle=\hat{U}_{\mathrm{np}}^{\sigma}\hat{S}(s_{\mathrm{np}})\vert0\rangle\vert g\rangle.
\end{equation}
Further, the average photon number in the ground state $\vert\varphi_{\mathrm{np}}^{G}\rangle$
can be calculated as 
\begin{equation}
\langle\hat{a}^{\dagger}\hat{a}\rangle_{\mathrm{np}}=\sinh^{2}s_{\mathrm{np}}+\frac{g^{2}}{\omega_{0}^{2}}-\frac{8g^{4}}{\omega_{0}^{4}}\mathrm{e}^{2s_{\mathrm{np}}},\label{eq:23-1}
\end{equation}
and the photon number variance can be obtained as 
\begin{equation}
\gamma'_{\mathrm{np}}\approx\frac{1}{2}\sinh^{2}(2s_{\mathrm{np}})+\frac{g^{2}}{\omega_{0}^{2}}\mathrm{e}^{-2s_{\mathrm{np}}}-\frac{8g^{4}\mathrm{e}^{4s_{\mathrm{np}}}}{\omega_{0}^{4}}.\label{eq:24-2}
\end{equation}
Then the LE can be calculated based on Eqs.~(\ref{eq:11}) and~(\ref{eq:24-2}).

\begin{figure}
\includegraphics[width=8.5cm]{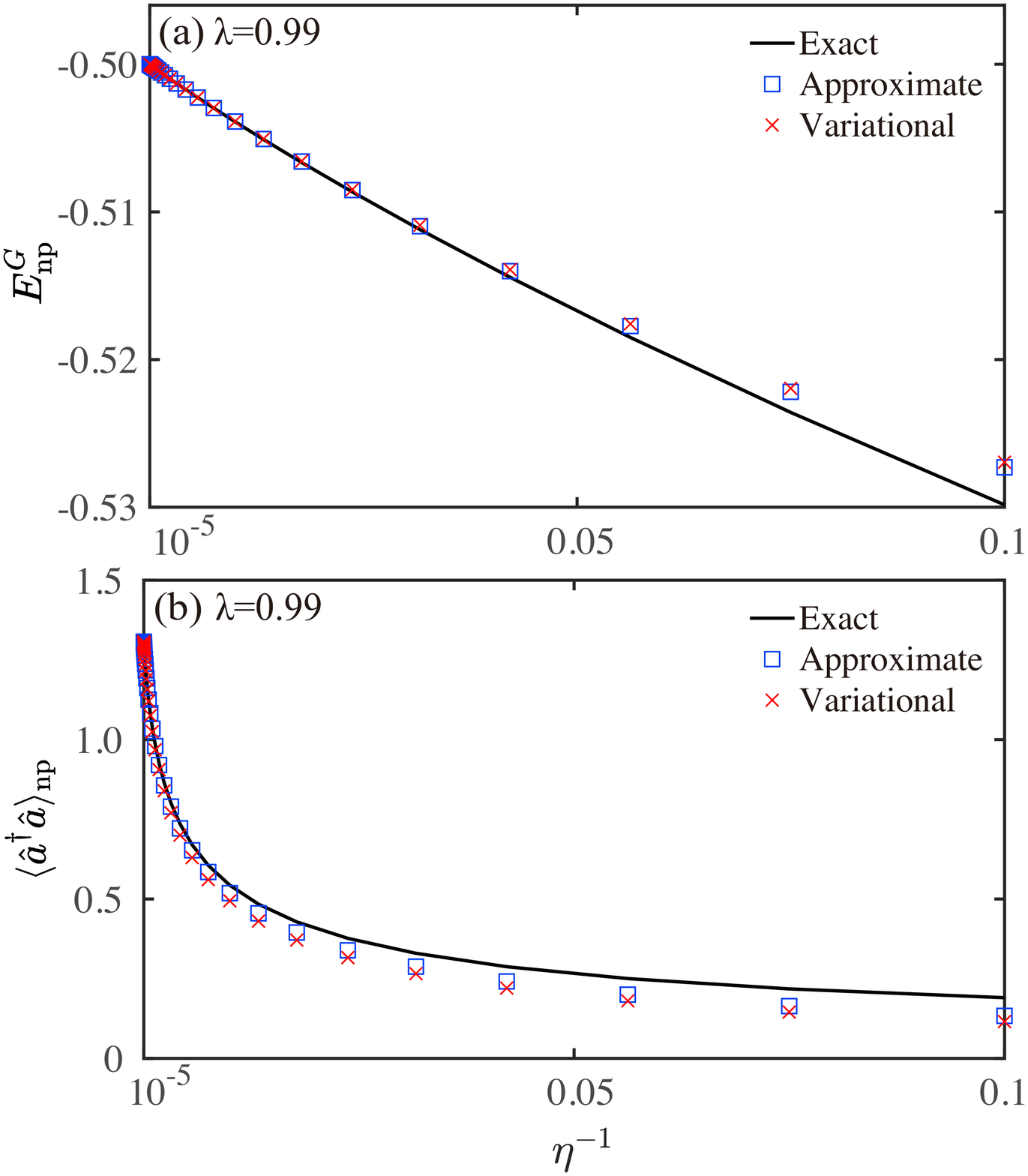} \caption{(Color online) (a) and (b) The ground-state energy and the average
photon number as functions of the frequency ratio $\eta$ in the normal
phase $(\lambda=0.99)$. Here, the exact and approximate results are
obtained based on the origin Hamiltonian~(\ref{eq:1}) (black solid
line) and the approximate Hamiltonian~(\ref{eq:19}) (squares). The
variational results are obtained based on Eqs.~(\ref{eq:19-1}),~(\ref{eq:21-1}),
and~(\ref{eq:23-1}) (crosses). }
\label{Fig:1} 
\end{figure}

We have used the variational method to solve the effective Hamiltonian
and obtained the photon number variance of the ground state in the
normal phase at a finite $\eta$. To check the validity of the effective
Hamiltonian and the approximate method, in Figs.~\ref{Fig:1}(a)
and~\ref{Fig:1}(b) we plot the ground-state energy and the average
photon number obtained by the variational and numerical methods, when
the QRM works in the normal phase $(\lambda=0.99)$. Meanwhile, we
present the exact result based on the origin Hamiltonian~(\ref{eq:1})
for reference. In Fig.~\ref{Fig:1}(a), the ground-state energy obtained
by these three methods in the normal phase are consistent with each
other in the large $\eta$ case. As the ratio $\eta$ decreases, the
deviation between the approximate result and the exact result in the
finite $\eta$ case becomes large. However, the variational result
agrees well with the numerical result of the effective Hamiltonian.
The average photon numbers obtained with these three methods match
well in the large $\eta$ case as shown in Fig.~\ref{Fig:1}(b).
In contrast, the difference between the variational result and the
numerical results increases with the decrease of $\eta$. The reason
for this difference is that the trial wave function only preserves
the low spin state, and the excite spin state is also important in
the finite $\eta$ case. Nevertheless, the variational method can
still catch the main physics.

\subsection{Photon number variance in the superradiance phase}

In this subsection, we calculate the photon number variance $\gamma$
in the ground state of the QRM working in the superradiance phase.
Here, we follow some derivations of the ground state of the QRM in
the superradiance phase given in Ref.~\cite{PhysRevLett.115.180404}
for keeping the completeness of this paper.

\subsubsection{The infinite $\eta$ case}

Physically, when the light-matter coupling strength increases to be
larger than the critical coupling strength, the coupled system will
acquire macroscopic excitations. Then the high-order terms which contain
the average photon number cannot be ignored. In this case, the approximate
Hamiltonian~(\ref{eq:13}) will not be valid as $\lambda>\lambda_{c}$.
To achieve the effective Hamiltonian in this case, we introduce a
displacement operator $\hat{D}(\alpha)=\exp[\alpha(a^{\dagger}-a)]$
to make a transformation upon Hamiltonian~(\ref{eq:1})~\cite{PhysRevLett.115.180404}
\begin{align}
\hat{\tilde{H}}_{\mathrm{Rabi}}(\alpha)= & \ \hat{D}^{\dagger}(\alpha)\hat{H}_{\mathrm{Rabi}}\hat{D}(\alpha)\nonumber \\
= & \ \omega_{c}(\hat{a}^{\dagger}+\alpha)(\hat{a}+\alpha)-g(\hat{a}+\hat{a}^{\dagger})\hat{\sigma}_{x}\nonumber \\
 & +\dfrac{\omega_{c}}{2}\hat{\sigma}_{z}-2g\alpha\hat{\sigma}_{x}.\label{eq:22}
\end{align}
Further, we introduce new spin eigenstates $\vert\tilde{e}\rangle$
and $\vert\tilde{g}\rangle$ of the atomic Hamiltonian $\omega_{c}\hat{\sigma}_{z}/2-2g\alpha\hat{\sigma}_{x}$.
In terms of the new spin states $\vert\tilde{e}\rangle=\cos\theta\vert e\rangle+\sin\theta\vert g\rangle$
and $\vert\tilde{g}\rangle=-\sin\theta\vert e\rangle+\cos\theta\vert g\rangle$
with $\tan(2\theta)=-4g\alpha/\omega_{0}$, the Pauli operators in
the new spin space can be defined by $\hat{\tau}_{0}=\vert\tilde{e}\rangle\langle\tilde{e}\vert+\vert\tilde{g}\rangle\langle\tilde{g}\vert$,
$\hat{\tau}_{x}=\vert\tilde{e}\rangle\langle\tilde{g}\vert+\vert\tilde{g}\rangle\langle\tilde{e}\vert$,
and $\hat{\tau}_{z}=\vert\tilde{e}\rangle\langle\tilde{e}\vert-\vert\tilde{g}\rangle\langle\tilde{g}\vert$.
Then Hamiltonian~(\ref{eq:22}) can be expressed as~\cite{PhysRevLett.115.180404}
\begin{align}
\hat{\tilde{H}}_{\mathrm{Rabi}}(\alpha)= & \ \omega_{c}\hat{a}^{\dagger}\hat{a}-g(\hat{a}+\hat{a}^{\dagger})\cos(2\theta)\hat{\tau}_{x}+\frac{\tilde{\omega}_{0}}{2}\hat{\tau}_{z}+\omega_{c}\alpha^{2}\nonumber \\
 & +[\omega_{c}\alpha\hat{\tau}_{0}-g\sin(2\theta)\hat{\tau}_{z}](\hat{a}+\hat{a}^{\dagger}).
\end{align}
To eliminate the block-diagonal perturbation term in the subspace
$\mathcal{H}_{\tilde{g}}$, we obtain, $\omega_{0}\alpha+g\sin(2\theta)=0$,
which leads to the displacement parameters~\cite{PhysRevLett.115.180404}
\begin{equation}
\alpha=\pm\alpha_{\lambda}=\pm\sqrt{\omega_{0}(\lambda^{4}-1)/(4\lambda^{2}\omega_{c})}.
\end{equation}
In the infinite-frequency limit, the term $2\omega_{c}\alpha(\hat{a}+\hat{a}^{\dagger})\vert\tilde{e}\rangle\langle\tilde{e}\vert$
can be ignored in the new low spin subspace, then the reformulated
Hamiltonian becomes~\cite{PhysRevLett.115.180404} 
\begin{equation}
\hat{\tilde{H}}_{\mathrm{Rabi}}(\pm\alpha_{\lambda})\approx\omega_{c}\hat{a}^{\dagger}\hat{a}+\frac{\tilde{\omega}_{0}}{2}\hat{\tau}_{z}^{\pm}-\tilde{g}(\hat{a}+\hat{a}^{\dagger})\hat{\tau}_{x}^{\pm}+\omega_{c}\alpha_{\lambda}^{2},\label{eq:25}
\end{equation}
where we introduce the parameters $\tilde{\omega}_{0}=\lambda^{2}\omega_{0}$
and $\tilde{g}=\sqrt{\omega_{c}\omega_{0}}/2\lambda$. The signs “$\pm$”
in $\hat{\tau}_{x,z}^{\pm}$ denote the direction of the displacement.
Note that the two different signs of the displacement parameter $\alpha$
indicate that the ground state exists twofold degeneracy in QRM~\cite{PhysRevLett.124.040404_2020}.
Hamiltonian~(\ref{eq:25}) has a similar structure as that of the
QRM, we could use a similar method to obtain the diagonalized Hamiltonian
by two unitary transformations as~\cite{PhysRevLett.115.180404}
\begin{align}
\hat{H}_{\mathrm{sp}}^{\mathrm{d}} & =\hat{S}^{\dagger}(r_{\mathrm{sp}})\hat{U}_{\mathrm{sp}}^{\dagger}\hat{\tilde{H}}_{\mathrm{Rabi}}\hat{U}_{\mathrm{sp}}\hat{S}(r_{\mathrm{sp}})\nonumber \\
 & =\hat{\epsilon}_{\mathrm{sp}}\hat{a}^{\dagger}\hat{a}+\hat{E}_{\mathrm{sp}}\label{eq:29-1}
\end{align}
where we introduce the conditional frequency $\hat{\epsilon}_{\mathrm{sp}}=\omega_{c}\sqrt{1+\lambda^{-4}\hat{\tau}_{z}^{\pm}}$
and the new spin-state dependent energy $\hat{E}_{\mathrm{sp}}=(\hat{\epsilon}_{\mathrm{sp}}-\omega_{c}+\omega_{0}\hat{\tau}_{z}^{\pm})/2+\omega_{c}\alpha_{\lambda}^{2}$.
The unitary transformation is defined as $\hat{U}_{\mathrm{sp}}=\exp[i(\tilde{g}/\tilde{\omega}_{0})(\hat{a}+\hat{a}^{\dagger})\hat{\tau}_{y}^{\pm}]$,
and the squeezing parameter here is $r_{\mathrm{sp}}=-\frac{1}{4}\ln(1+\lambda^{-4}\hat{\tau}_{z}^{\pm})$.
The ground state of the diagonalized Hamiltonian~(\ref{eq:29-1})
in the superradiance phase is $\left|0\right\rangle \left|\tilde{g}^{\pm}\right\rangle $.
Similarly, to keep the Hamiltonian $\hat{H}_{\mathrm{sp}}^{\mathrm{d}}$
in the low spin subspace to be Hermitian, the coupling strength $g$
should be larger than $\sqrt{\omega_{0}\omega_{c}}/2$ such that $\lambda>\lambda_{c}=1$,
which defines the parameter space of the \textit{superradiance phase}.
The ground state of the QRM in the superradiance phase can be expressed
as~\cite{PhysRevLett.115.180404} 
\begin{equation}
\vert\psi_{\mathrm{sp}}^{G}(r_{\mathrm{sp}})\rangle_{\pm}=\hat{D}(\pm\alpha_{\lambda})\hat{U}_{\mathrm{sp}}\hat{S}(r_{\mathrm{sp}})\vert0\rangle\vert\tilde{g}^{\pm}\rangle,\label{eq:28}
\end{equation}
where the transformation operators have defined before. In the ground
state $\vert\psi_{\mathrm{sp}}^{G}\rangle_{\pm}$, the photon number
variance in the superradiance phase can be obtained as 
\begin{equation}
\gamma_{\mathrm{sp}}=\frac{1}{2}\sinh^{2}(2r_{\mathrm{sp}})+\alpha_{\lambda}^{2}\mathrm{e}^{2r_{\mathrm{sp}}}+\frac{\tilde{g}^{2}}{\tilde{\omega}_{0}^{2}}\mathrm{e}^{-2r_{\mathrm{sp}}}.\label{eq:29}
\end{equation}
In terms of Eqs.~(\ref{eq:11}) and~(\ref{eq:29}), the analytical
result of the LE in the superradiance phase can be obtained.

\subsubsection{The finite $\eta$ case}

To beyond the infinite-frequency case, below we calculate the photon
number $\gamma_{\mathrm{sp}}^{\prime}$ by the variational method.
To this end, we perform a unitary transfomation with the transformation
operator
\begin{equation}
\hat{U}_{\mathrm{sp}}^{\sigma}=\exp\left\{ i\left[\frac{\tilde{g}}{\tilde{\omega}_{0}}(\hat{a}+\hat{a}^{\dagger})-\frac{4\tilde{g}^{3}}{3\tilde{\omega}_{0}^{3}}(\hat{a}+\hat{a}^{\dagger})^{3}\right]\hat{\tau}_{y}^{\pm}\right\} 
\end{equation}
to Hamiltonian~(\ref{eq:25}) and projecting the transformed Hamiltonian
into the low spin subspace, then the effective Hamiltonian becomes~\cite{PhysRevLett.115.180404}
\begin{equation}
\hat{H}_{\mathrm{sp}}^{\sigma}=\omega_{c}\hat{a}^{\dagger}\hat{a}-\frac{\tilde{g}^{2}}{\tilde{\omega}_{0}}(\hat{a}+\hat{a}^{\dagger})^{2}+\frac{\tilde{g}^{4}}{\tilde{\omega}_{0}^{3}}(\hat{a}+\hat{a}^{\dagger})^{4}-\frac{\tilde{\omega}_{0}}{2}+\frac{\tilde{g}^{2}\omega_{c}}{\tilde{\omega}_{0}^{3}}+\omega_{c}\alpha_{\lambda}^{2}.\label{eq:31}
\end{equation}

Similar to the treatment in the normal-phase case, the trial wave
function in the superradiance phase is assumed as $\vert\Psi_{\mathrm{sp}}^{G}(s_{\mathrm{sp}})\rangle=\hat{S}(s_{\mathrm{sp}})\vert0\rangle\vert\tilde{g}^{\pm}\rangle$.
The corresponding ground state energy can be obtained as 
\begin{align}
E_{\mathrm{sp}}^{G}(s_{\mathrm{sp}})= & \ \omega_{c}\sinh^{2}s_{\mathrm{sp}}-\frac{\omega_{c}}{4\lambda^{4}}\mathrm{e}^{2s_{\mathrm{sp}}}+\frac{3\omega_{c}^{2}}{16\tilde{\omega}_{0}\lambda^{8}}\mathrm{e}^{4s_{\mathrm{sp}}}\nonumber \\
 & -\frac{\tilde{\omega}_{0}}{2}+\frac{\omega_{c}^{2}}{4\tilde{\omega}_{0}\lambda^{4}}+\omega_{c}\alpha_{g}^{2},\label{eq:35}
\end{align}
where the parameter $s_{\mathrm{sp}}$ is determined by the zero point
of the first-order derivative, 
\begin{equation}
\frac{\mathrm{d}E_{\mathrm{sp}}^{G}(s_{\mathrm{sp}})}{\mathrm{d}s_{\mathrm{sp}}}=\ \frac{\omega_{c}}{2\mathrm{e}^{2s_{\mathrm{np}}}}\left[\frac{3\mathrm{e}^{6s_{\mathrm{sp}}}}{2\eta\lambda^{10}}+(1-\lambda^{-4})\mathrm{e}^{4s_{\mathrm{sp}}}-1\right]=0.\label{eq:33}
\end{equation}
The only physical solution of Eq.~(\ref{eq:33}) is given by 
\begin{equation}
s_{\mathrm{sp}}=\frac{1}{2}\ln\left\{ \mathrm{Re}\left[\frac{\sqrt[3]{B}}{9}-\frac{2(\lambda^{4}-1)}{9\lambda^{-6}\eta^{-1}}+\frac{4(\lambda^{4}-1)^{2}}{9\sqrt[3]{B}\lambda^{-12}\eta^{-2}}\right]\right\} ,\label{eq:38}
\end{equation}
where we introduce 
\begin{align}
B= & \ 9\sqrt{3}\sqrt{243\lambda^{20}\eta^{2}+(1-\lambda^{12}+3\lambda^{8}-3\lambda^{4})16\lambda^{28}\eta^{4}}\nonumber \\
 & +243\lambda^{10}\eta+(-\lambda^{12}+3\lambda^{8}-3\lambda^{4}+1)8\lambda^{18}\eta^{3}.
\end{align}
The ground state of QRM in this case can be expressed as 
\begin{equation}
\vert\varphi_{\mathrm{sp}}^{G}(s_{\mathrm{sp}})\rangle=\hat{D}(\pm\alpha_{\lambda})\hat{U}_{\mathrm{sp}}^{\sigma}\hat{S}(s_{\mathrm{sp}})\vert0\rangle\vert\tilde{g}^{\pm}\rangle.\label{eq:39-1}
\end{equation}
Based on the ground state~(\ref{eq:39-1}), the average photon number
in the superradiance phase can be calculated as 
\begin{equation}
\langle\hat{a}^{\dagger}\hat{a}\rangle_{\mathrm{sp}}=\sinh^{2}s_{\mathrm{sp}}+\frac{\tilde{g}^{2}}{\tilde{\omega}_{0}^{2}}-\frac{8\tilde{g}^{4}}{3\tilde{\omega}_{0}^{4}}\mathrm{e}^{2s_{\mathrm{sp}}}+\alpha_{g}^{2},\label{eq:39}
\end{equation}
and the photon number variance can be obtained 
\begin{equation}
\gamma'_{\mathrm{sp}}\approx\frac{1}{2}\sinh^{2}(2s_{\mathrm{sp}})+\frac{\tilde{g}^{2}}{\tilde{\omega}_{0}^{2}}\mathrm{e}^{-2s_{\mathrm{sp}}}+\left(\alpha_{g}^{2}-\frac{8\tilde{g}^{4}\mathrm{e}^{2s_{\mathrm{sp}}}}{3\tilde{\omega}_{0}^{4}}\right)\mathrm{e}^{2s_{\mathrm{sp}}}.
\end{equation}

\begin{figure}
\includegraphics[width=8.5cm]{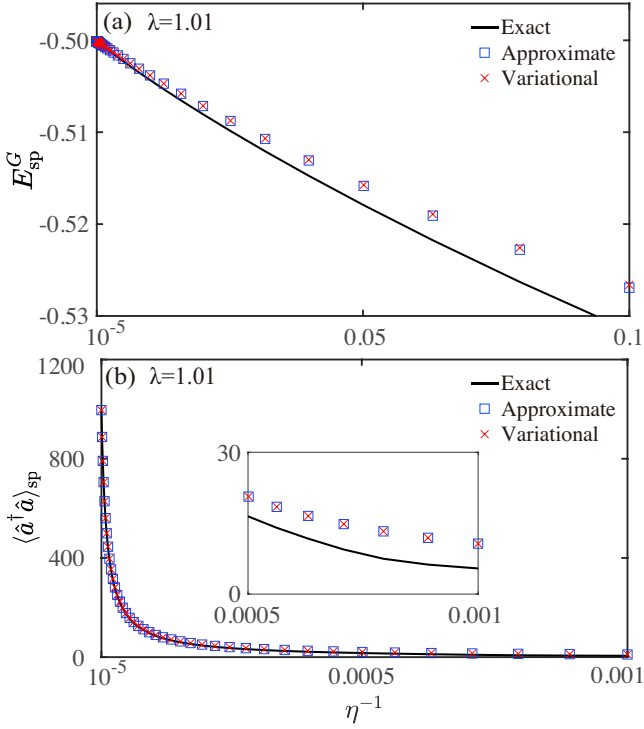} \caption{(Color online) (a) and (b) The ground-state energy and the average
photon number as functions of the frequency ratio $\eta$ in the superradiance
phase $(\lambda=1.01)$. Here, the exact and approximate results are
obtained based on the displaced Hamiltonian~(\ref{eq:22}) (black
solid line) and the approximate Hamiltonian~(\ref{eq:31}) (squares),
respectively. The variational results are obtained based on Eqs.~(\ref{eq:35}),~(\ref{eq:38}),
and~(\ref{eq:39}) (crosses). }
\label{Fig:2} 
\end{figure}

In order to examine the validity of the effective Hamiltonian~(\ref{eq:31})
and the variational method in the superradiance phase, in Figs.~\ref{Fig:2}(a)
and~\ref{Fig:2}(b) we compare the variational result with the numerical
result through solving the effective Hamiltonian~(\ref{eq:31}) and
the displaced Hamiltonian (\ref{eq:22}), respectively. In Fig.~\ref{Fig:2}(a),
we plot the ground-state energy of the QRM solved based on the exact
displaced Hamiltonian, the approximate Hamiltonian, and the variational
method. Here, we find that the results based on these three methods
match better as the frequency ratio $\eta$ increases. For the average
photon number as plotted in Fig.~\ref{Fig:2}(b), we observe that
its average value is very large, which corresponds to the characteristics
of the system in the superradiance phase. We also see that the variational
results match the numerical results well in the superradiance phase
(inset), which indicates the validity of the variational method.

\begin{figure}
\center \includegraphics[width=8cm]{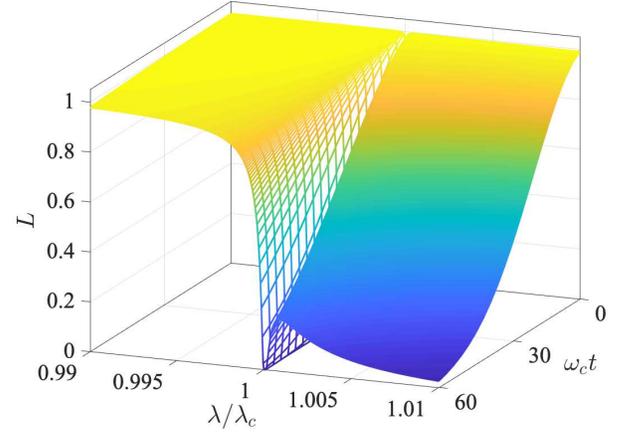} \caption{(Color online) The LE of the auxiliary atom as a function of the scaled
coupling strength $\lambda/\lambda_{c}$ and the scaled evolution
time $\omega_{c}t$ in both the normal phase ($\lambda<\lambda_{c}$)
and the superradiance phase ($\lambda>\lambda_{c}$). Here, we take
$\eta=5000$ and $\chi=g_{s}^{2}/\Delta_{s}=0.001\omega_{c}$.}
\label{Fig:3} 
\end{figure}

By far, we have already calculated the photon number variance of the
QRM in both the normal phase and the superradiance phase. In the next
section, we will study the LE of the auxiliary atom corresponding
to the QRM in the normal and superradiance phases.

\section{The LOSCHMIDT ECHO}

\label{section4}

In this section, we study how to exhibit the critical dynamics of
the QRM by checking the LE of the auxiliary atom when the system goes
across the critical point from the normal phase to the superradiance
phase. In the short-time limit, the LE has been simplified to Eq.~(\ref{eq:11}).
Here, the LE can reflect the main character of the QPT by the quantum
decoherence of the auxiliary atom. To show the dependence of the LE
on the criticality, in Fig.~\ref{Fig:3} we plot the LE versus the
dimensionless coupling strength $\lambda$ and the evolution time
$t$ based on Eqs.~(\ref{eq:11}), (\ref{eq:16}), and (\ref{eq:29})
in the normal and superradiance phases. Figure~\ref{Fig:3} shows
that, in the vicinity of the critical point, the LE experiences a
sharp change within a small range of $\lambda/\lambda_{c}$. In the
normal phase, the LE decays sharply to zero as the dimensionless coupling
strength $\lambda$ approaches the critical point $\lambda_{c}$.
In the superradiance phase, the LE decays faster as the parameter
$\lambda$ increases far away from the critical point $\lambda_{c}$,
and reaches the minimal value at a large coupling strength. The rapid
change indicates that the coherence of the auxiliary atom is supersensitive
to a perturbation inflicted on the QRM near the critical point. We
can measure the QPT of QRM based on the supersensitive coherence of
the auxiliary atom in QRM. In addition, the coherence of the auxiliary
atom near the critical point decreases to zero sharply with time at
the critical point $\lambda_{c}$. During this process, the detected
atom evolves from a pure state to a mixed one.

\begin{figure}[t]
\center \includegraphics[width=8.5cm]{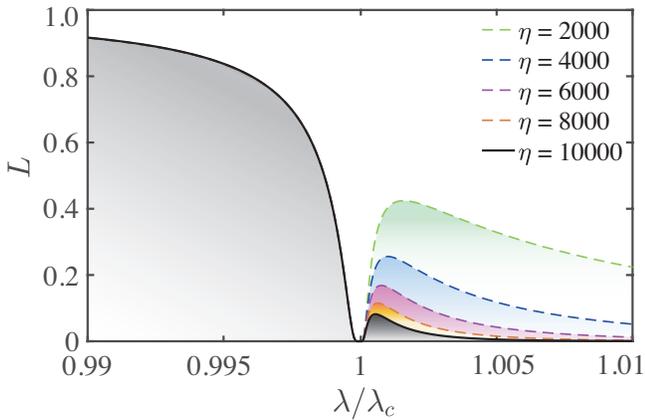} \caption{(Color online) The LE of the auxiliary atom versus the scaled coupling
strength $\lambda/\lambda_{c}$ at $\omega_{c}t=60$ when $\eta$
takes different values: $\eta=2000$ (green dashed line), $\eta=4000$
(blue dashed line), $\eta=6000$ (purple dashed line), $\eta=8000$
(orange dashed line), and $\eta=10000$ (black solid line). Other
parameters used are the same as those given in Fig.~\ref{Fig:3}.}
\label{Fig:4} 
\end{figure}

In our previous discussions, we have discussed the ground states of
the QRM in both infinite and finite $\eta$ cases. To know the influence
of the ratio $\eta$ on the LE, below we study the dependence of the
LE on the parameter $\lambda$ at different values of $\eta$. In
Fig.~\ref{Fig:4}, the LE is plotted as a function of $\lambda$
at a fixed time when the ratio $\eta$ takes different values: $\eta=2000,4000,6000,8000,$
and $10000$. Here, we can see that, in the normal phase, the LE decays
from a finite value to zero when the scaled coupling strength $\lambda/\lambda_{c}$
increases approach to one. In particular, the LE is independent of
the ratio $\eta$ in the normal phase. In the superradiance phase,
with the increase of the ratio $\lambda/\lambda_{c}$, the LE experiences
an increase from zero to peak values and then decays to zero. Different
from the normal phase, the revival peak value of the LE is smaller
for a larger value of $\eta$.

\begin{figure}[tbh]
\center \includegraphics[width=8.5cm]{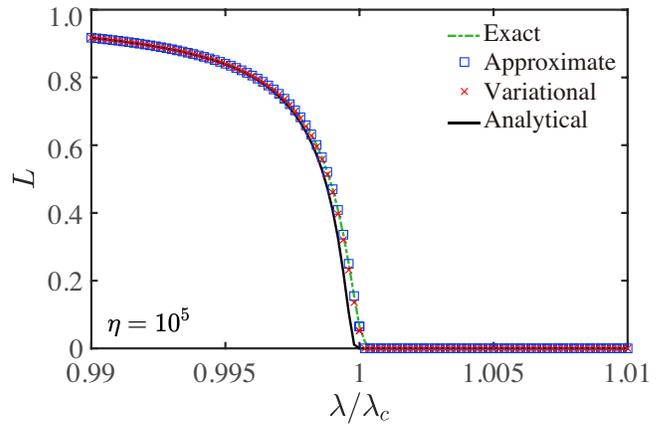} \caption{(Color online) The LE of the auxiliary atom versus the scaled coupling
strength $\lambda/\lambda_{c}$ at $\omega_{c}t=60$. These curves
are plotted with different methods: the numerical result based on
the effective Hamiltonians given by Eqs.~(\ref{eq:19}) and~(\ref{eq:31})
(squares), the numerical result based on the original Rabi Hamiltonian
given by Eq.~(\ref{eq:1}) in the normal phase and Eq.~(\ref{eq:22})
in the superradiance phase (green dashed line), and the variational
method (crosses) for $\eta=10^{5}$. We also present the analytical
result in the infinite $\eta$ case for reference (black solid line).
Other parameters used are the same as those given in Fig.~\ref{Fig:3}.}
\label{Fig:5} 
\end{figure}

It should be pointed out that, though our analytical discussions are
valid for the infinite $\eta$ case, the dynamic sensitivity of the
quantum criticality also exists in the large finite $\eta$ case.
To show this point, in Fig.~\ref{Fig:5} we plot the LE of the auxiliary
atom as a function of $\lambda/\lambda_{c}$ at a large finite $\eta=10^{5}$.
For comparison, here we plot the LE using three different methods.
We numerically solve the dynamics governed by the effective Hamiltonians
given in Eqs.~(\ref{eq:19}) and~(\ref{eq:31}). The exact numerical
results are based on the original Rabi Hamiltonian~(\ref{eq:1})
in the normal phase and the displaced Hamiltonian~(\ref{eq:22})
in the superradiance phase. We also plot the LE based on the variational
method. In addition, we present the analytical result in the infinite
$\eta$ case for reference~\cite{PhysRevLett.115.180404}. Here,
we can see that, the quantum criticality exists in the large finite
$\eta$ case, and that the results obtained with three methods are
in consistent with each other. In the normal phase, the LE decays
from a finite value to zero, and there is no obvious revival in the
superradiance phase.

\section{SUMMARY}

\label{section5}

In summary, we have studied the dynamic sensitivity of quantum phase
transition in the QRM by checking the LE of an auxiliary atom, which
is far-off-resonantly coupled to the cavity field of the QRM. In the
vicinity of the critical point, the LE displays a sudden decay, which
is associated with the quantum decoherence of the auxiliary atom.
We have checked quantum criticality of the QRM in both the infinite
and finite $\eta$ cases. The analytical results in the infinite $\eta$
case clearly indicate the dynamic sensitivity in this model. Moreover,
when the ratio $\eta$ between the frequency of the two-level atom
and frequency of the cavity field is finite but large, i.e., in the
finite $\eta$ case, the effective Hamiltonian can also reflect the
quantum criticality in the QRM. Our proposal provides a simple scheme
for observation of quantum criticality in the QRM by checking the
quantum coherence of an atomic sensor. 
\begin{acknowledgments}
J.-Q.L. is supported in part by National Natural Science Foundation
of China (Grants No.~11822501, No.~11774087, and No.~11935006)
and Hunan Science and Technology Plan Project (Grant No.~2017XK2018).
J.-F.H. is supported in part by the National Natural Science Foundation
of China (Grant No.~12075083), Scientific Research Fund of Hunan
Provincial Education Department (Grant No.~18A007), and Natural Science
Foundation of Hunan Province, China (Grant No.~2020JJ5345). Q.-T.X.
is supported in part by National Natural Science Foundation of China
(Grants No.~11965011). 
\end{acknowledgments}

\providecommand{\noopsort}[1]{}\providecommand{\singleletter}[1]{#1}%

\end{document}